# The role of ferroelectric-ferromagnetic layers on the properties of superlattice-based multiferroics

P. Murugavel, M.P. Singh, W. Prellier[1], B. Mercey, Ch. Simon and B. Raveau

Laboratoire CRISMAT, CNRS UMR 6508, ENSICAEN,

6 Bd du Maréchal Juin, F-14050 Caen Cedex, France.

**Abstract**

A series of superlattices and trilayers composed of ferromagnetic and ferroelectric or paraelectric layers were grown on (100) $SrTiO_3$ by the pulsed laser deposition technique. Their structural and magneto-electric properties were examined. The superlattices made of ferromagnetic $Pr_{0.85}Ca_{0.15}MnO_3$ (PCMO) and a ferroelectric, namely $Ba_{0.6}Sr_{0.4}TiO_3$ (BST) or $BaTiO_3$, showed enhanced magnetoresistance (MR) at high applied magnetic field, whereas such enhancement was absent in $Pr_{0.85}Ca_{0.15}MnO_3/SrTiO_3$ superlattices, which clearly demonstrates the preponderant role of the ferroelectric layers in this enhanced MR. Furthermore, the absence of enhanced MR in trilayers of PCMO/BST indicates that the magneto-electric coupling which is responsible for MR in these systems is stronger in multilayers than in their trilayer counterparts.

**P.A.C.S: 81.15 Fg, 75.47.Gk, 75.70.Cn**

---

[1] prellier@ensicaen.fr

## 1. Introduction

In recent years, there has been growing interest in tailoring materials for the coexistence of two or more properties, which are usually absent in the parent forms[1], and the associated coupling between them allows the additional degree of freedom in the designing of devices, such as transducers, actuators and information storage devices[2-4]. In the multiferroics, ferromagnetic and ferroelectric ordering exists simultaneously. Hence, one expects the direct coupling between the magnetic and dielectric properties and their control by the application of magnetic and/or electric fields[4]. However, there are only few materials which behave as multiferroics[4-5]. Recently, efforts have been made to fabricate the artificial layers and tailor their structures for the suitability of multiferroics[6-9]. These materials are in the form of either composites, superlattices, or multilayers. Interestingly, the material made in the form of superlattices, whose structure consists of alternating ferroelectric and ferromagnetic layers, yielded unusual electrical and magnetic transport properties that cannot be obtained in either of their constituents. Recently, we reported the transport properties of superlattices made by fixing the ferromagnetic $Pr_{0.85}Ca_{0.15}MnO_3$ (PCMO) layer constant and varying the thickness of ferroelectric $Ba_{0.6}Sr_{0.4}TiO_3$ (BST) layer, in which an unusual enhanced MR was observed. Indeed a negative MR of as high as 35% at 100 K was demonstrated in the superlattice consisting of higher ferroelectric layer thickness (*i.e.* 9 unit cells of BST)[9]. Such enhancement in MR was attributed to the ferroelectric spacer layer and the associated magnetoelectric coupling with the ferromagnetic layer[9].

In this article, we have studied the magneto-transport properties of the PCMO/BST superlattices by keeping the BST layer constant to 9 unit cells and varying only the thickness of the PCMO layer. To understand the role of interfaces, we have also investigated the properties of trilayers based on PCMO/BST/PCMO and BST/PCMO/BST with an equivalent thickness of their counterparts. Furthermore, to clarify the role of ferromagnetic or ferroelectric layers in the associated enhanced MR effect, we have studied the magneto-transport properties of several other superlattices composed of ferromagnetic-ferroelectric such as PCMO/BaTiO$_3$ (BTO), La$_{0.7}$Sr$_{0.3}$MnO$_3$ (LSMO)/BTO, and ferromagnetic/paraelectric PCMO/SrTiO$_3$ (STO), and our results are reported in this article. The enhanced MR in the superlattices of PCMO/BST and PCMO/BTO, and its absence in the PCMO/STO, LSMO/BTO superlattices and trilayers suggest the importance of both the ferroelectric layer and the magnetoelectric couplings.

## 2. Experimental

Superlattices of (PCMO$_N$/BST$_9$)$_{25}$, (PCMO$_{10}$/STO$_9$)$_{25}$, (PCMO$_{10}$/BTO$_9$)$_{25}$, and (LSMO$_{10}$/BTO$_9$)$_{25}$ along with the trilayers of PCMO/BST/PCMO, and BST/PCMO/BST were prepared on (100)-SrTiO$_3$ (STO) substrates (where $N$ is the number of PCMO unit cells and varies from 1 to 10) by multi-target pulsed-laser deposition method. The targets were prepared by conventional solid state route using powders of Pr$_2$O$_6$, CaCO$_3$, MnO$_2$, BaCO$_3$, SrCO$_3$, and TiO$_2$. All samples were deposited at 720$^\circ$C at 100 mTorr of oxygen pressure and followed by the deposition; substrate was cooled down to room temperature at the rate of 13°C/min under 300 Torr of oxygen pressure. A series of PCMO/BST superlattices were made by keeping the BST thickness constant at 9 unit cells (u.c.) and varied the PCMO layer from 3 to 10 unit cells. The 9 u.c. was chosen for BST, because it gave the higher MR effect[9]. The structure of the samples was examined by x-ray diffraction (XRD) using Seifert 3000P diffractometer (Cu K$\alpha_1$, $\lambda$=1.5406 Å) and transmission electron microscopy (TEM). The resistance ($R$) was measured by applying the current perpendicular-to-the-plane and magnetization ($M$) was measured as a function of temperature ($T$) and magnetic field ($H$) using a superconducting quantum interference device magnetometer (SQUID). The magnetoresistance (MR) is calculated as $MR=100\times[R(H)-R(0)]/R(0)$ where R(H) is the resistance at the magnetic field H.

## 3. Results and discussion

### 3.1. Structural properties

The XRD scans of $PCMO_N/BST_9$ superlattices recorded around the (002) fundamental peak along with the simulation of the superlattice structure using *DIFFaX* program[10] are shown in Fig. 1a. It is found that the experimentally measured peaks are in good agreement with the simulated one in terms of position and intensity. Furthermore the presence of higher order satellite peaks (the number $i$ corresponds to the $i^{th}$ satellite peak in Fig.1a), which arising from the chemical modulation of multilayer structure, clearly reveals that the layers were grown hetero-epitaxially with sharp interfaces. To examine the structure of $PCMO_N/BST$ superlattices in greater depth, high resolution-TEM (HRTEM) images were recorded on various samples. A typical HRTEM image of $PCMO_9/BST_9$ supelattice is shown in Fig. 1b. It confirms the crystalline nature of the films. A contrast consisting of a regular array of bright spots is observed in the enlarged image of the interface (inset of Fig.1b), which clearly demonstrates the epitaxial nature of the films and corroborates our XRD findings. Surprisingly, the interface between PCMO and BST is not clearly visible, which is basically due to the following reasons. First, the defocus values and second, the Z-values are almost equal for BST and PCMO[11,12]. Detailed TEM and XRD studies show no evidence for the presence of any impurity phases in the samples confirming the good quality of the samples.

### 3.2. Magnetic properties

The magnetic properties of the $PCMO_N/BST_9$ superlattices are shown in Fig. 2. As seen from the field-dependence of the magnetization (normalized to per unit cell of PCMO layer) plotted in Fig. 2a, the saturation magnetization is not observed (even at 5 kOe). This could be due to a spin canting effect near the interfaces of the superlattices[13]. We did not see any significant change in the magnetization curve vs. temperatures (the shape of the hysteresis curves are similar) with the increase in the number of PCMO unit cells in the superlattices. In addition, all superlattices are ferromagnetic with a slight change in the Curie temperature ($T_C$). For example, the $T_C$ of the superlattice shows only a slight increase from 100 to 120 K with increase in *N*, most probably due to a much better lattice mismatch resulting in a coherently strain layer[14]. The coercive field of the samples is increasing with a progressive increase in PCMO thickness and reaches to its parent PCMO film. Variations in the magnetization of the films have also been observed (Fig 2a). Usually the variation in the magnetic property of the superlattices arises from the variation of the strain in them. Moreover, in the present case variation is also partly coming due to increase in the total magnetic volume of the film, i.e. the increase of the PCMO thickness. Furthermore the variation in magnetic transition temperature ($T_C$) can be attributed to the variation in the strain of the film. In addition, this is due to the fact that with the progressive increase in PCMO layer thickness, the growth is more coherent[14]. In order to understand the role of the different layers, we have measured the transport properties.

### 3.3. Role of ferromagnetic/ferroelectric layers

The corresponding MR of the PCMO$_N$/BST$_9$, measured at 100 K, in current-perpendicular-to plane configuration using LaNiO$_3$ as electrode[9] is shown in Fig. 2b. First, the maximum MR is observed for the sample PCMO$_{10}$/BST$_9$ (the MR of parent PCMO films is only 4 %) and second, the MR value decreases with the decrease of the thickness of the PCMO layer. This is not surprising since the magnetic layer is PCMO. One possible explanation of the enhanced MR, is the change in resistance associated with the negative differential resistance of the PCMO layer[10] with electric charges developing at the interface (due to the ferroelectric BST layer and the magnetoelectric coupling in the superlattice[9]). However, to clarify the role of ferroelectric BST layer, we have measured the MR for several other superlattices which are made by changing the type of the different spacer layers, *i.e.* ferroelectric and paraelectric. Thus, PCMO/BST, PCMO/BTO, PCMO/STO, and LSMO/BTO superlattices were made and the MR results are shown in Fig.3. Two important points are observed. First, at 100 K (near the T$_C$ of the PCMO), MR values were found to be 35 %, 18.5 %, 4.5 %, and 4 % for PCMO$_{10}$/BST$_9$, PCMO$_{10}$/BTO$_9$, PCMO$_{10}$/STO$_9$, and PCMO samples, respectively. Second, only the superlattices made of PCMO and the ferroelectric layers (BST and BTO) show enhanced MR compared to the PCMO/STO superlattice whose MR value is nearly same as that of PCMO film itself (4 %). Also, note that the Curie temperature (T$_C$) of the superlattices are nearly identical (not shown), indicating that the difference in MR are not due to the difference in T$_C$. Thus, the above result clearly shows that the ferroelectric layer is essential for the enhancement in MR observed in the PCMO/BST film. This variation reveals that there is possible magneto-electric coupling in the ferromagnetic-ferroelectric superlattices, which will be absent in their ferromagnetic-paraelectric counterparts. Furthermore, presence of MR in all the superlattices, also exhibits the coherent spin transport in these films. Moreover, the smaller value of MR observed in PCMO/BTO superlattices in contrast with the PCMO/BST may be attributed to the difference in their interfacial strains. The lattice mismatch between PCMO and BST is 2.3 %, whereas it is 3.5 % between PCMO and BTO suggesting that the lattice strains at interface plays an important role on the transport properties of the superlattice.

In order to ascertain the role of the magnetic layer, we have also prepared superlattices made of ferroelectric BTO and another manganite oxide, namely LSMO, which is a ferromagnetic metal with a $T_C$ close to room temperature. A series of superlattices of $LSMO_{10}/BTO_N$ were made in the same way as PCMO/BTO superlattice and their transport properties measured at 250 K, below the Curie temperature $T_C$. The $T_C$ of all the superlattices is close to 300 K, *i.e.*, slightly smaller than the parent LSMO film grown under identical conditions, as a result of spin canting at the interface[13] and/or interfacial effects[15]. As a representative example, in Fig. 3b MR versus magnetic field is plotted for $LSMO_{10}/BTO_9$ measured at 250 K and for LSMO measured at 250 and 300 K. Unlike PCMO/BTO, the enhancement is not observed in LSMO/BTO superlattices, which suggests that the magnetoelectric coupling is stronger in PCMO-based superlattices than that of the LSMO-based superlattices. Further investigations are underway to extract the role of magnetic oxides on the enhancement of MR.

Furthermore, to examine the role of interface, we have also fabricated the trilayers composed of PCMO/BST/PCMO and BST/PCMO/BST. These films were also grown under identical conditions on STO with an equivalent thickness of their superlattice counterparts, *e.g.,* for $PCMO_{10}/BTO_9$ superlattice, trilayers with 250 unit cells of PCMO and 225 unit cells of BST were grown. Magnetic measurements performed on several trilayers reveal that there is no significant variation in the transition temperatures and coercive fields. MR measurements were also carried out on these samples. The observed MR in trilayers was on the order of only ~ 2 %. The absence of large MR in the trilayers is an indirect proof that the interface, which is the same for the trilayers and the superlattices, does not play a significant role in the observed enhanced MR in the superlattices. Thus it clearly suggests that the magnetoelectric coupling is stronger in superlattices than that of the trilayers.

**Conclusion**

We have successfully prepared the superlattice-based multiferroics of $(Pr_{0.85}Ca_{0.15}MnO_3)_N/(Ba_{0.6}Sr_{0.4}TiO_3)_9$ along with $(Pr_{0.85}Ca_{0.15}MnO_3)_N/(BaTiO_3)_9$, $(Pr_{0.85}Ca_{0.15}MnO_3)_N/(SrTiO_3)_9$ and $(La_{0.7}Sr_{0.3}MnO_3)_N/(BaTiO_3)_9$ to clarify the role of ferromagnetic manganite oxide and the ferroelectric layer upon their transport properties. The enhancement in magnetoresistance is observed only for the superlattices composed of $Pr_{0.85}Ca_{0.15}MnO_3$ and ferroelectric layer and the maximum magnetoresistance is observed for the films with higher $Pr_{0.85}Ca_{0.15}MnO_3$ and $Ba_{0.6}Sr_{0.4}TiO_3$ thickness. The magnetoelectric coupling in the superlattices is considered as the main cause for the observed enhancement in magnetoresistance.

**Acknowledgements**

We also would like to thank Dr. H. Eng and Dr. P. Padhan for fruitful discussions.

This work has been carried out in the frame of the Work Package "*New Architectures for Passive Electronics*" of the European Network of Excellence "*Functionalized Advanced Materials Engineering of Hybrids and Ceramics*" FAME (FP6-500159-1) supported by the European Community, and by Centre National de la Recherche Scientifique (CNRS). One of the authors (P.M) acknowledges the Ministère de la Jeunesse et de l'Education Nationale for his fellowship (2003/87).

Figure Captions:

Figure 1: (a) Measured and simulated XRD scan recorded around the (002) reflection of SrTiO$_3$ substrates for various superlattices (PCMO$_N$/BST$_9$)$_{25}$ (*N*=3-10). The symbol *i* indicates the order of the satellite peak. (b) High resolution TEM cross-section image of a (PCMO$_N$/BST$_9$)$_{25}$ superlattice. The inset shows the magnified segment of the interface (noted as a white arrow) between the substrate and the superlattice.

Figure 2 : (a) Magnetization loop of (PCMO$_N$/BST$_9$)$_{25}$ superlattice measured at 10 K. The inset shows the evolution of the Curie temperature (T$_C$) and the coercive field (H$_C$) as a function of the spacer layer thickness (*N*) for (PCMO$_N$/BST$_9$)$_{25}$ superlattices. (b) Corresponding MR recorded at 100 K as a function of magnetic field.

Figure 3 : (a) Evolution of MR recorded at 100 K as a function of magnetic field measured for (PCMO$_{10}$/BST$_9$)$_{25}$, (PCMO$_{10}$/BTO$_9$)$_{25}$, (PCMO$_{10}$/STO$_9$)$_{25}$, and PCMO samples. (b) MR measured at 250 and 300 K for LSMO and (LSMO$_{10}$/BTO$_9$)$_{25}$ samples.

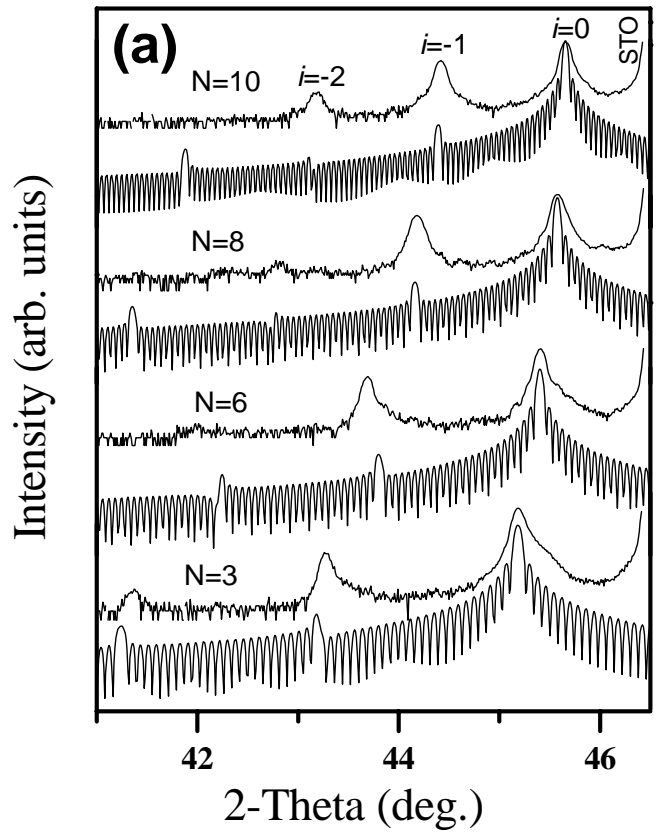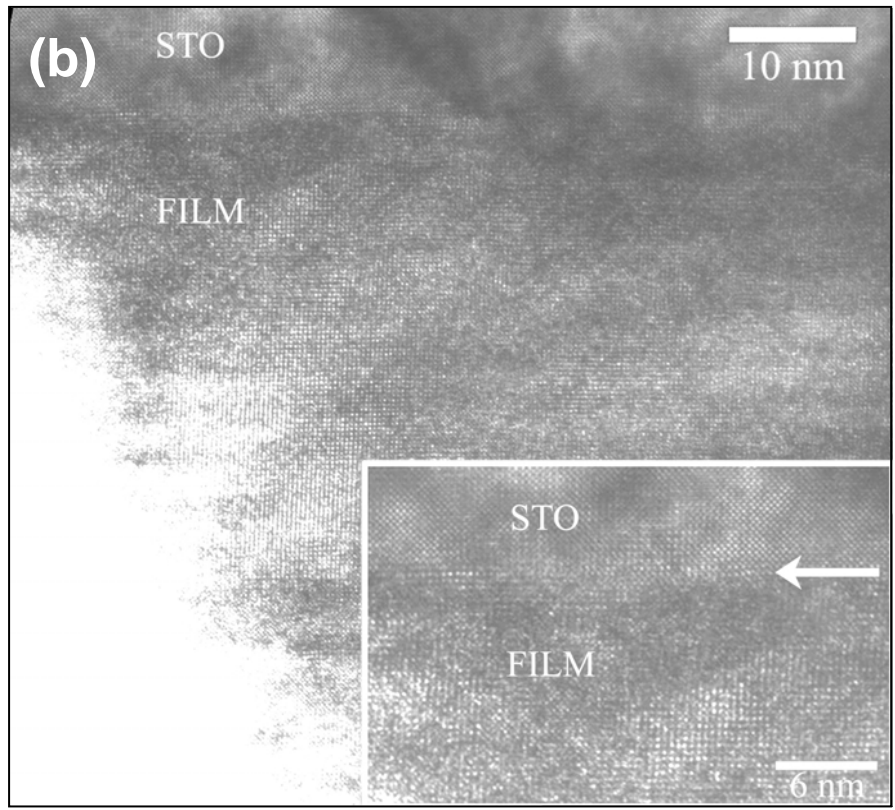

P. Murugavel et al., Figure 1

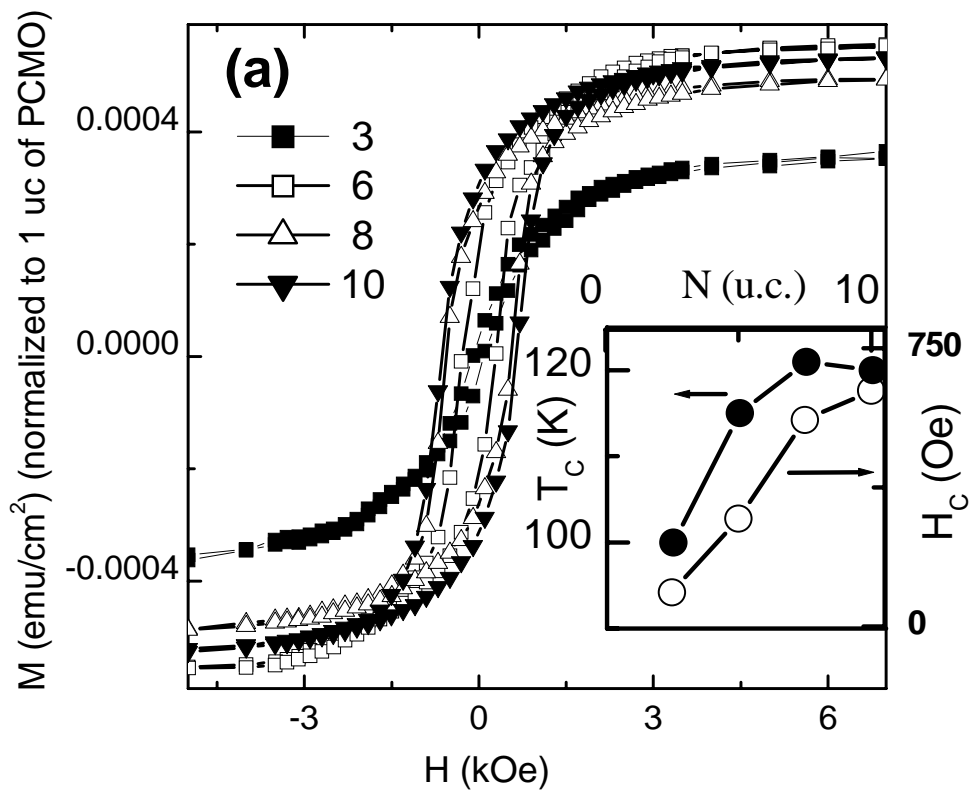
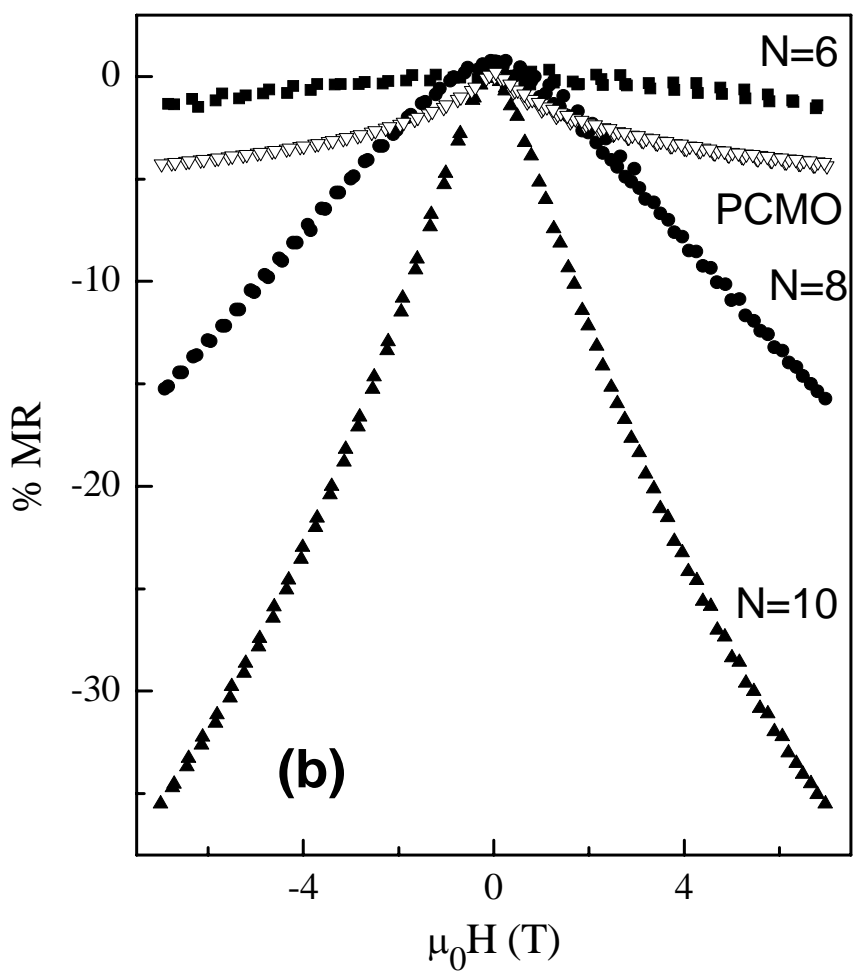

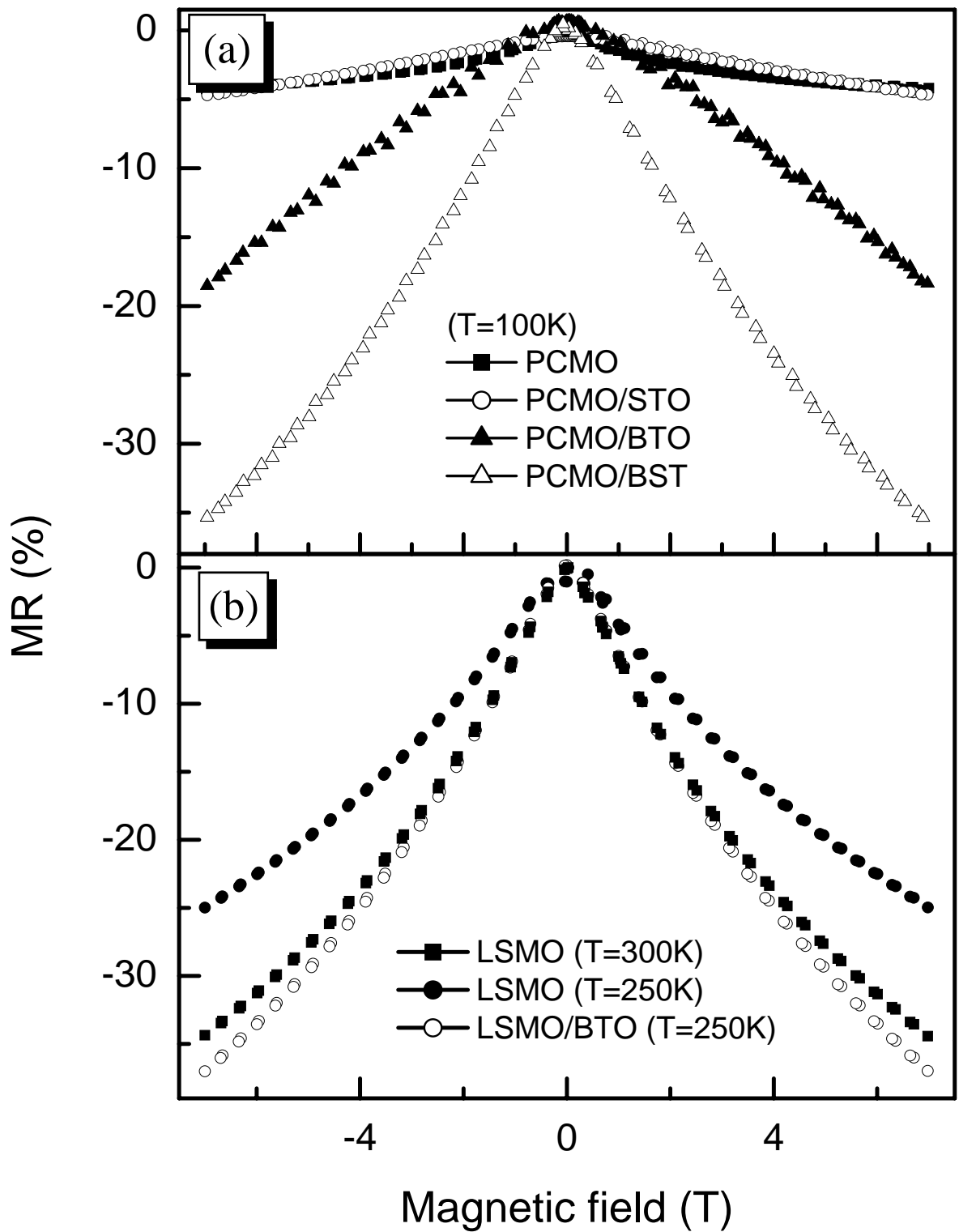

Figure 3
P. Murugavel et al.